\def\ra{\rangle}
\def\la{\langle}
\def\be{\begin{equation}}
\def\ee{\end{equation}}
\def\ba{\begin{array}}
\def\ea{\end{array}}

\documentclass[aps, pre, preprint, showpacs, amsmath, amssymb, amsfonts, preprintnumbers]
{revtex4}
\usepackage{epsfig, graphicx}
\usepackage{amsmath}
\input amssym.def

\begin{document}
\title{Tight upper bound for the maximal quantum value of the Svetlichny operators}
\author{Ming Li$^{1,2}$}
\author{Shuqian Shen$^{1}$}
\author{Naihuan Jing$^{3, 2}$}
\author{Shao-Ming Fei$^{2,4}$}
\author{Xianqing Li-Jost$^{2}$}

\affiliation{$^1$ College of the Science, China University of
Petroleum, 266580 Qingdao, China\\
$^2$ Max-Planck-Institute for Mathematics in the Sciences, 04103
Leipzig, Germany\\
$^3$ Department of Mathematics, Shanghai University, 200444, Shanghai, China\\
$^4$ School of Mathematical Sciences, Capital Normal University, Beijing 100048, China}

\begin{abstract}
It is a challenging task to detect genuine multipartite nolocality (GMNL). In this paper, the problem is considered via computing the maximal quantum value of Svetlichny operators for three-qubit systems and a tight upper bound is obtained. The constraints on the quantum states for the tightness of the bound are also presented. The approach enables us to give the necessary and sufficient conditions of violating the Svetlichny inequalities (SI) for several quantum states, including the white and color noised GHZ states. The relation between the genuine multipartite entanglement concurrence and the maximal quantum value of the Svetlichny operators for mixed GHZ class states is also discussed. As the SI is useful for the investigation on GMNL, our results give an effective and operational method to detect the GMNL for three-qubit mixed states.
\end{abstract}

\smallskip

\pacs{03.67.-a, 02.20.Hj, 03.65.-w} \maketitle

\section{Introduction}
Quantum nonlocality, which is incompatible
with the local hidden variable (LHV) theory, can be revealed via violations of various Bell
inequalities \cite{bell,chsh,vbell1,vbell2,vbell3,chenjingling,liprl,yu}. It has been recognized that quantum nonlocality is not only a puzzling aspect of nature, but also an important resource for quantum information processing, such as building
quantum protocols to decrease communication complexity \cite{dcc,dcc1}
and providing secure quantum communication \cite{scc1,scc2}.

For the multipartite case, quantum nonlocality displays much richer and more
complex structures than the bipartite case \cite{dgmnl}. One can distinguish qualitatively different kinds
of nonlocality. In this manuscript, we consider the tripartite case.
If Alice, Bob and Charlie perform measurement $X$, $Y$ and $Z$ on the three subsystems, respectively, with
outcomes $a$, $b$ and $c$, and the probability correlations $P(abc|XYZ)$ among the measurement outcomes
can be written as
\be
P(abc|XYZ) =\sum_{\lambda}q_{\lambda}P_{\lambda}(a|X) P_{\lambda}(b|Y) P_{\lambda}(c|Z),
\ee
where $0\leq q_{\lambda}\leq 1$ and $\sum_{\lambda}q_{\lambda}=1$, then the state measured is called three local. Otherwise the state is non-three local. Svetlichny \cite{svetlichny} pointed out that some correlations can be written in the hybrid local-nonlocal (or bi-LHV) form,
\begin{eqnarray}\label{bilhv}
&&P(abc|XYZ)\nonumber\\
&=&\sum_{\lambda}q_{\lambda} P_{\lambda}(ab|XY) P_{\lambda}(c|Z)
+\sum_{\mu}q_{\mu} P_{\mu}(ac|XZ) P_{\mu}(b|Y)
+\sum_{\nu}q_{\nu} P_{\nu} (bc|YZ)P_{\nu} (a|X),
\end{eqnarray}
where $0\leq q_{\lambda}, q_{\mu}, q_{\nu} \leq 1$ and $\sum_{\lambda}q_{\lambda}+\sum_{\mu}q_{\mu}+\sum_{\nu}q_{\nu}=1$.
Correlations without the above form are called genuine tripartite nonlocal (or genuine three-way nonlocal).

Svetlichny has formulated a hybrid nonlocal-local
realism based inequality \cite{svetlichny}: a stronger kind of inequality
for a three-qubit system where two of the qubits are assumed to be non-locally correlated, but they are locally
correlated to the third, with an ensemble average over all such possible combinations. The violation of the Svetlichny inequality (SI) is a
signature of genuine tripartite nonlocality. We refer to \cite{brunner,def1,def2,def3,def4} for more details.

The question studied in this paper is that for an arbitrary (pure or mixed) three-qubit state, how one checks whether the SI is violated
or not. The authors in \cite{indian} has considered this problem for pure GHZ class states and W class states, and explicit formula for the maximal values of the Svetlichny operator
over all the measurement observables are obtained. In \cite{piano} the authors have provided analytical and numerical prescriptions for detecting the maximum violation of the SI for pure and mixed Gaussian states of continuous variable systems.

In this paper we consider the computation of the maximal quantum value of the Svetlichny operators for any three-qubit systems.
We present a tight upper bound for the maximal quantum value. We also provide the constraints on the quantum state for the tightness of the bound. Moreover, the sufficient and necessary condition of violating the SI for several quantum states is given, including the white and color noised GHZ states.  We also discuss the relation between the genuine tripartite entanglement concurrence and the maximal quantum value of the Svetlichny operators for mixed GHZ class states. As the SI is powerful to investigate GMNL, our results give an effective and operational way to detect the GMNL for three-qubit mixed states.

\section{The Svetlichny inequality and the maximal violation}

We start with a short review of the Svetlichny inequality \cite{svetlichny}.
The Svetlichny operators $\mathcal{S}$ are defined as follows:
\begin{equation}\label{siop}
 \mathcal{S}=A\otimes\left[(B+B')\otimes C+(B-B')\otimes C'\right]+A'\otimes\left[(B-B')\otimes C-(B+B')\otimes C'\right],
\end{equation}
where $A,A',B,B',C$ and $C'$ are observables of the form $G=\vec{g}\cdot\vec{\sigma}=\sum_kg_{k}\sigma_{k},G\in\{A,A',B,B',C,C'\}$ and
$\vec{g}\in \{\vec{a},\vec{a}',\vec{b},\vec{b}',\vec{c},\vec{c}'\}$ correspondingly, $\sigma_{k}$ $(k=1,2,3)$ are the Pauli matrices, $\vec{\sigma}=(\sigma_{1},\sigma_{2},\sigma_{3})$,
$\vec{g}=(g_{1},g_{2},g_{3})$ is a 3-dimensional real unit vector.
For any 3-qubit state $|\Psi\rangle$ with the bi-LHV form (\ref{bilhv}), the mean value of the Svetlichny operators is bounded as follows \cite{svetlichny}:
\begin{equation}
\mid\langle\Psi|\mathcal{S}|\Psi\rangle\mid\leq4.
\end{equation}

{\bf{Theorem:}}
For any three-qubit quantum state $\rho$, the maximum quantum value $Q(\mathcal{S})$ of the Svetlichny operator $\mathcal{S}$ defined in (\ref{siop}) satisfies
\begin{eqnarray}
Q(\mathcal{S})\equiv\max\mid \langle\mathcal{S}\rangle_{\rho}\mid\nonumber\leq 4\lambda_{1},
\end{eqnarray}
where $\langle\mathcal{S}\rangle_{\rho}=Tr(\mathcal{S}\,\rho)$, $\lambda_{1}$ is the maximum singular value of the matrix $M=(M_{j,ik})$, 
with $M_{ijk}=tr\left[\rho(\sigma_{i}\otimes\sigma_{j}\otimes\sigma_{k})\right], i,j,k=1,2,3$.

To prepare for the argument 
we first give the following result.

{\bf{Lemma:}}\label{T:rect4} Let $A$ be a rectangular matrix of size $m\times n$. For any vectors $\vec{x}\in\mathbb R^m$ and $\vec{y}\in\mathbb R^n$  we have that
\begin{equation}\label{lm}
|\vec{x}^TA\vec{y}|\leq \lambda_{max}|\vec{x}||\vec{y}|,
\end{equation}
where $\lambda_{max}$ is the largest singular value of the matrix $A$. The equality holds when
$\vec{x}$ and $\vec{y}$ are the corresponding singular vectors of $A$ with respect to $\lambda_{max}$.

{\bf Proof}. By the singular value decomposition, there exist two unitary matrices $U$ and $V$ such that $A=U^T\Sigma V$, where $\Sigma$ has only nonzero entries along the diagonal.
Therefore, we may assume that $A=\Sigma$ and consider only the following form,
$G(\vec{x}, \vec{y})=\sum_ia_ix_iy_i$,
where $a_1\geq a_2\geq \cdots\geq a_n>0$. Using the Cauchy-Schwarz inequality for the inner product $\langle \vec{x}, \vec{y}\rangle:=G(\vec{x}, \vec{y})$, we have that
\begin{align*}
|G(\vec{x}, \vec{y})|&\leq G(\vec{x}, \vec{x})^{1/2}G(\vec{y}, \vec{y})^{1/2}\\
&=(\sum_i a_ix_i^2)^{1/2}(\sum_i a_iy_i^2)^{1/2}\\
&\leq a_1(\sum_i x_i^2)^{1/2}(\sum_i y_i^2)^{1/2},
\end{align*}
where $a_1$ corresponds to $\lambda_{max}$ in (\ref{lm}).
\hfill \rule{1ex}{1ex}

{\bf Proof of the theorem.} By definition we have that
\begin{eqnarray*}
  \mathcal{S} &=& A\otimes\left[(B+B')\otimes C+(B-B')\otimes C'\right]+A'\otimes\left[(B-B')\otimes C-(B+B')\otimes C'\right]\\
              &=& \sum_{i,j,k}\left[a_{i}(b_{j}+b'_{j})c_{k}+a_{i}(b_{j}-b'_{j})c'_{k}+a'_{i}(b_{j}-b'_{j})c_{k}-a'_{i}(b_{j}+b'_{j})c'_{k}\right]
                  \sigma_{i}\otimes\sigma_{j}\otimes\sigma_{k}.
\end{eqnarray*}
Then
\begin{eqnarray*}
\langle \mathcal{S}\rangle_\rho&=& \sum_{i,j,k}\left[a_{i}(b_{k}+b'_{k})c_{j}+a_{i}(b_{k}-b'_{k})c'_{j}+a'_{i}(b_{k}-b'_{k})c_{j}-a'_{i}(b_{k}+b'_{k})c'_{j}\right]
    tr\left[\rho(\sigma_{i}\otimes\sigma_{k}\otimes\sigma_{j})\right]\\
&=& \sum_{i,j,k}\left[a_{i}(b_{k}+b'_{k})c_{j}+a_{i}(b_{k}-b'_{k})c'_{j}+a'_{i}(b_{k}
     -b'_{k})c_{j}-a'_{i}(b_{k}+b'_{k})c'_{j}\right]M_{ij, k} \\
&=& (\vec{a}\otimes \vec{c}-\vec{a}'\otimes \vec{c}')^TM(\vec{b}+\vec{b}')+ (\vec{a}\otimes \vec{c}'+\vec{a}'\otimes \vec{c})^TM(\vec{b}-\vec{b}').
\end{eqnarray*}

Now denote by $\theta_a$ (resp. $\theta_b$, $\theta_c$) the angle between $\vec{a}$ and $\vec{a}'$ (resp. $\vec{b}$ and $\vec{b}'$, $\vec{c}$ and $\vec{c}'$), we then have
\begin{align*}
|\vec{b}+ \vec{b}'|^2&=2+\langle \vec{b}, \vec{b}'\rangle=2+2\cos\theta_b=4\cos^2\frac{\theta_b}2,\\
|\vec{b}- \vec{b}'|^2&=2+\langle \vec{b}, \vec{b}'\rangle=2-2\cos\theta_b=4\sin^2\frac{\theta_b}2,\\
|\vec{a}\otimes \vec{c}-\vec{a}'\otimes \vec{c}'|^2&=2-2\langle \vec{a}, \vec{a}'\rangle\langle \vec{c}, \vec{c}'\rangle=2-2\cos\theta_a\cos\theta_c,\\
|\vec{a}\otimes \vec{c}'+\vec{a}'\otimes \vec{c}|^2&=2+2\langle \vec{a}, \vec{a}'\rangle\langle \vec{c}, \vec{c}'\rangle=2+2\cos\theta_a\cos\theta_c.
\end{align*}
Let $\theta_{ac}$ be the principal angle such that
$\cos\theta_a\cos\theta_c=\cos\theta_{ac}$. Therefore
\begin{align*}
|\vec{a}\otimes \vec{c}-\vec{a}'\otimes \vec{c}'|^2&=4\sin^2\frac{\theta_{ac}}2,\\
|\vec{a}\otimes \vec{c}'+\vec{a}'\otimes \vec{c}|^2&=4\cos^2\frac{\theta_{ac}}2.
\end{align*}

It follows from the Lemma (\ref{lm}) and trigonometric identities that
\begin{align}\label{proof}
|\langle \mathcal{S}\rangle_\rho| &\leq
\lambda_1\left(|\vec{a}\otimes \vec{c}-\vec{a}'\otimes \vec{c}'||\vec{b}+\vec{b}'|+|\vec{a}\otimes \vec{c}'+\vec{a}'\otimes \vec{c}||\vec{b}-\vec{b}'| \right)\nonumber\\
&=4\lambda_1\left(|\sin\frac{\theta_{ac}}2\cos\frac{\theta_b}2|+|\cos\frac{\theta_{ac}}2\sin\frac{\theta_b}2|\right)\nonumber\\
&=4\lambda_1|\sin\left(\frac{\theta_{ac}\pm\theta_b}2\right)|\nonumber\\
&\leq 4\lambda_1,
\end{align}
which proves the inequality. \hfill \rule{1ex}{1ex}

Let's look at when the equality holds. Actually, to saturate the upper bound in the Theorem, one can always select $\theta_{ac}\pm\theta_b=\pi$ or $-\pi$
by setting proper measurement directions of $B$ and $B'$ such that the last inequality in (\ref{proof}) becomes an equality. Then from the
Lemma, we have that the first inequality in (\ref{proof}) saturates if the degeneracy of $\lambda_1$ is more than one, and corresponding to $\lambda_1$ there are two 9-dimensional singular vectors taking the form of $\vec{a}\otimes \vec{c}-\vec{a}'\otimes \vec{c}'$ and $\vec{a}\otimes \vec{c}'+\vec{a}'\otimes \vec{c}$, respectively. The next examples show that the upper bound is tight.

\textbf{Example 1:}
We consider the mixture of the white noise and the three-qubit GHZ-class states, which is given by
\begin{equation}\label{ghz}
\rho=p|\psi_{gs}\rangle\langle\psi_{gs}|+\frac{1-p}{8}I,
\end{equation}
where $I$ is identity matrix,
$|\psi_{gs}\rangle=\cos\theta|000\rangle+\sin\theta|11\rangle(\cos\theta_{3}|0\rangle+\sin\theta_{3}|1\rangle)$,
and
$0\leq p\leq 1$.
The matrix $M$ is of the form,
\begin{equation}
M =p\left({\begin{array}{*{20}{c}}
a&0&b&0&-a&0&0&0&0\\
0&-a&0&-a&0&-b&0&0&0\\
0&0&0&0&0&0&c&0&d
\end{array}}\right),
\end{equation}
where $a=2\cos\theta \sin\theta \sin\theta_{3}, b=2\cos\theta \cos\theta_{3}\sin\theta, c=2\cos\theta_{3}\sin^{2}\theta \sin\theta_{3}$ and
$d=\cos^{2}\theta+\sin^{2}\theta \cos^{2}2\theta_{3}$.

The singular values of the matrix $M$ are
$p\cdot |\sin2\theta|\sqrt{1+\sin^{2}\theta_{3}}, p\cdot |\sin2\theta|\sqrt{1+\sin^{2}\theta_{3}},$ and $p\cdot \sqrt{1-\sin^{2}2\theta \sin^{2}\theta_{3}}$.
Hence the upper bound of the maximal value of the Svetlichny operators is
\begin{eqnarray}
Q(\mathcal{S})\equiv\max\mid \langle\mathcal{S}\rangle_{\rho}\mid \leq 4\lambda_{1}=4p\cdot\mathop{max}_{\theta,\theta_{3}}
\{|\sin2\theta|\sqrt{1+\sin^{2}\theta_{3}},\sqrt{1-\sin^{2}2\theta \sin^{2}\theta_{3}}\}.
\end{eqnarray}

One finds that $p\cdot \sqrt{1-\sin^{2}2\theta \sin^{2}\theta_{3}}$ is always less than one. Thus to violate the SI, we can just take
$\lambda_1=p\cdot |\sin2\theta|\sqrt{1+\sin^{2}\theta_{3}}$.
To prove that the upper bound is saturated for the mixed state $\rho$ in (\ref{ghz}), we set, for convenience,
$$x=-\frac{\sin2\theta\sin\theta_3}{\sin^22\theta(1+\sin^2\theta_3)}, ~~~~~ y=-\frac{\sin2\theta\cos\theta_3}{\sin^22\theta(1+\sin^2\theta_3)}.
$$
The singular vectors corresponding to $\lambda_{1,2}$ can be selected as
$\left(
\begin{array}{ccc}
0,
x,
0,
x,
0,
y,
0,
0,
0
\end{array}
\right)^T
$ and $
\left(
\begin{array}{ccc}
-x ,
0 ,
-y ,
0 ,
x ,
0 ,
0 ,
0 ,
0
\end{array}
\right)^T$,
which can be directly decomposed as
\begin{equation*}
\left(
\begin{array}{ccc}
1,
0,
0
\end{array}
\right)^T
\otimes
\left(
\begin{array}{ccc}
0,
x,
0
\end{array}
\right)^T+\left(
\begin{array}{ccc}
0,
-1,
0
\end{array}
\right)^T
\otimes
\left(
\begin{array}{ccc}
-x,
0,
-y
\end{array}
\right)^T
\end{equation*}
and \begin{equation*}
\left(
\begin{array}{ccc}
1,
0,
0
\end{array}
\right)^T
\otimes
\left(
\begin{array}{ccc}
-x,
0,
y
\end{array}
\right)^T-\left(
\begin{array}{ccc}
0,
-1,
0
\end{array}
\right)^T
\otimes
\left(
\begin{array}{ccc}
0,
x,
0
\end{array}
\right)^T.
\end{equation*}
Then we can set
\begin{equation*}\vec{a}=
\left(
\begin{array}{ccc}
1,
0,
0
\end{array}
\right)^T,~ \vec{a}'=
\left(
\begin{array}{ccc}
0,
-1,
0
\end{array}
\right)^T,~ \vec{c}=
\frac{1}{\sqrt{x^2+y^2}}\left(
\begin{array}{ccc}
-x,
0,
-y
\end{array}
\right)^T,~ \vec{c}'=
\left(
\begin{array}{ccc}
0,
1,
0
\end{array}
\right)^T,
\end{equation*}
and set $\vec{b}$ and $\vec{b}'$ to be unit vectors s.t. $\theta_{ac}+\theta_b=2\pi$.
With the above settings one can find that each of the inequalities in the proof of the theorem becomes equal,
which means that the upper bound is saturated for $\rho$.

Note that if we set $p=1$, then the maximal value is in accordance with the main result in \cite{indian}.
However, the theorem in this manuscript fits for arbitrary mixed three-qubit systems.
By the analysis above, one obtains that $\rho$ in (\ref{ghz}) will violate the SI if and only if $ {1}/{|\sin2\theta|\sqrt{1+\sin^{2}\theta_{3}}}<p\leq 1$. If we set $\theta_3=\frac{\pi}{2}$ and $\theta=\frac{\pi}{4}$, the state $\rho$ in (\ref{ghz}) is exactly the $\rho_{GHZ}$: the mixture of the GHZ state $|GHZ\ra=\frac{1}{\sqrt{2}}(|000\ra+|111\ra)$ and white noise. From our Theorem we have that $\rho_{GHZ}$ will violate the SI if and only if $0.707107<p\leq1$, which is in accordance with the result in \cite{dgmnl} from numerical optimization. By \cite{hashemi} we have that $\rho_{GHZ}$ will be genuine multipartite entangled if and only if $0.428571<p\leq1$, see Fig. 1.

\begin{figure}[h]
\begin{center}
\resizebox{14cm}{!}{\includegraphics{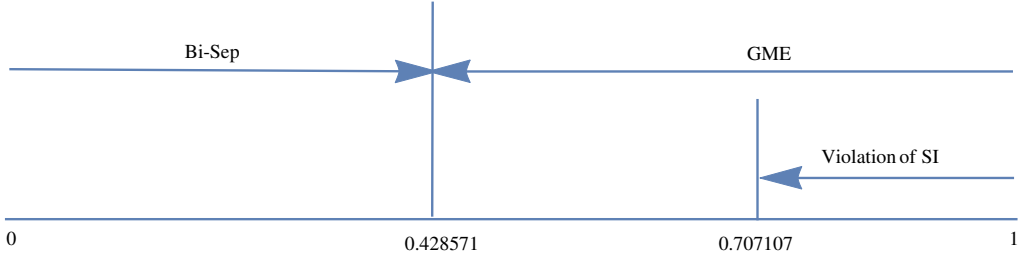}}
\end{center}
\caption{Consider the mixture of GHZ state and white noise: $\rho_{GHZ}=p|GHZ\ra\la GHZ|+\frac{1-p}{8}I$, where $|GHZ\ra=\frac{1}{\sqrt{2}}(|000\ra+|111\ra)$. By \cite{hashemi} we have that $\rho_{GHZ}$ will be genuine multipartite entangled if and only if $0.428571<p\leq1$, while by our theorem one obtains that $\rho_{GHZ}$ will violate the SI if and only if $0.707107<p\leq1$.\label{fig1}}
\end{figure}

\textbf{Example 2:} Consider the quantum state $\sigma_A(\rho)$ presented in \cite{prl030404}. Set $N=3$ and $d=2$. Then $\sigma_A(\rho)$ is a three-qubit state,
\be
\sigma_A(\rho)=p|GHZ\ra\la GHZ|+\frac{1-p}{4}I_2\otimes \tilde{I},
\ee
where $I_2$ stands for the $2\times 2$ identity matrix and $\tilde{I}=\mathrm{diag}(1, 0, 0, 1)$. 
As analyzed in \cite{prl030404}, $\sigma_A(\rho)$ admits bi-local hidden models for $0\leq p\leq0.416667$, and is a genuine multipartite entangled state for  $\frac{1}{3}< p\leq 1$.
By our theorem, one gets
\begin{equation}
M =\left({\begin{array}{*{20}{c}}
p&0&0&0&-p&0&0&0&0\\
0&-p&0&-p&0&0&0&0&0\\
0&0&0&0&0&0&0&0&0
\end{array}}\right),
\end{equation}
and can further check that the upper bound is $4\sqrt{2}p$ and is saturated, which means that $\sigma_A(\rho)$ will be GMNL for $0.707107<p\leq1$, as shown in Fig. 2.

\begin{figure}[h]
\begin{center}
\resizebox{14cm}{!}{\includegraphics{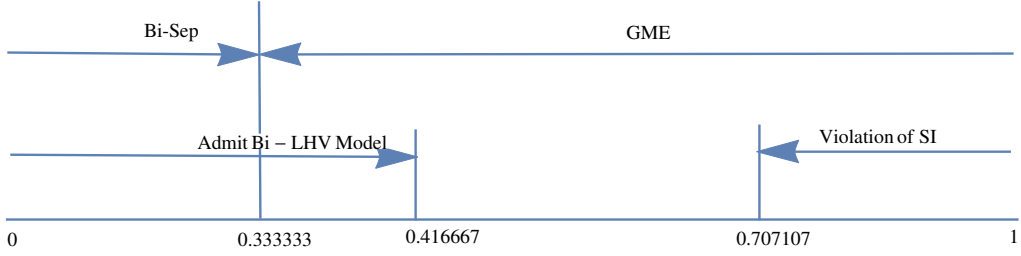}}
\end{center}
\caption{Consider the mixture of GHZ state and color noise: $\sigma_A(\rho)=p|GHZ\ra\la GHZ|+\frac{1-p}{4}I_2\otimes \tilde{I}$. By \cite{prl030404}, one has that $\sigma_A(\rho)$ admit bi-local hidden models for $0\leq p\leq0.416667$, and are genuine multipartite entangled for $\frac{1}{3}< p\leq 1$, while by our theorem one obtains that $\sigma_A(\rho)$ will be GMNL for $0.707107<p\leq1$.\label{fig2}}
\end{figure}

In the following we consider the relation between the genuine multipartite entanglement(GME) concurrence and the SI bound obtained above.
The GME concurrence is proved to be a well defined measure \cite{ma1,ma2}.
For a tripartite pure state $|\psi\ra$, the GME concurrence is defined by
\begin{eqnarray*}
C_{GME}(|\psi\ra)=\sqrt{\min\{1-tr(\rho_1^2),1-tr(\rho_2^2),1-tr(\rho_3^2)\}},
\end{eqnarray*}
where $\rho_i$ is the reduced matrix for the $i$th subsystem.
For mixed states $\rho$, the GME concurrence is then defined by the convex roof,
\begin{eqnarray}
C_{GME}(\rho)=\min\sum_{\{p_{\alpha},|\psi_{\alpha}\ra\}}p_{\alpha}C_{GME}(|\psi_{\alpha}\ra).
\end{eqnarray}
The minimum is taken over all pure ensemble decompositions of $\rho$.

In \cite{submitted} we have already found a lower bound for GME concurrence for three-qubit quantum systems. For a three-qubit state $\rho$, the GME concurrence satisfies the following inequality \cite{submitted},
\begin{eqnarray}\label{lb}
C_{GME}(\rho)\geq\sqrt{\frac{1}{8}||M||_{HS}^2}-\frac{1}{2},
\end{eqnarray}
where $||M||_{HS}$ stands for the Hilbert-Schmidt norm or 2-norm of matrix $M$.
For the state (\ref{ghz}), we have
\begin{eqnarray}\label{14}
C_{GME}(\rho)&\geq&\sqrt{\frac{1}{8}||M||_{HS}^2}-\frac{1}{2}
=\sqrt{\frac{1}{8}(\lambda_1^2+\lambda_2^2+\lambda_3^2)}-\frac{1}{2}\nonumber\\
&=&\sqrt{\frac{1}{64}Q^2(\mathcal{S})+\frac{1}{8}(p^2-p^2\sin^2 2\theta \sin^{2}\theta_{3})}-\frac{1}{2}\nonumber\\
&\geq&\frac{1}{8}Q(\mathcal{S})-\frac{1}{2}
\end{eqnarray}
which presents an explicit relation satisfied by the GME concurrence and the maximal value of the SI operators.

Based on the tight upper bound of the maximal quantum value of the Svetlichny operators, we have presented a lower bound of GME concurrence for mixed GHZ class states.
(\ref{14}) implies that if the maximal value of the SI operators is greater than 4, then the GME concurrence is greater than zero.
Namely, as long as the state (\ref{ghz}) does not admit the bi-LHV form, it must be genuine multipartite entangled.
Moreover, the lower bound for the GME concurrence is experimentally friendly as the mean value of the SI operator can measured experimentally.
Hence (\ref{14}) can also serve as an effective experimentally-friendly criterion for detecting genuine multipartite entanglement.

\section{Conclusions and Discussions}\label{sec4}

We have presented a quantitative analysis of
the genuine tripartite nonlocality for any three-qubit quantum systems via effective computation of the maximal quantum value of the Svetlichny operators. Our method provides a tight upper bound for the maximal quantum value. The tightness of the upper bound is investigated through several noisy quantum states. Our result works not only for pure states but also for mixed states.
Since the SI is powerful in detecting GMNL, our results give an effective and operational way to investigate the GMNL for three-qubit mixed states. We have also discussed the relation between the GME concurrence and the maximal quantum value of the Svetlichny operators.
Based on the tight upper bound of the maximal quantum value of the Svetlichny operators, we have presented a lower bound for GME concurrence for mixed GHZ class states, which serves as an effective and experimentally-friendly criteria for detecting GME.
The method presented in this manuscript can also be used in computing the maximal violations of other tripartite or multipartite Bell-type inequalities such as that in \cite{dgmnl, bancal} or the Svetlichny inequalities for the continuous quantum systems \cite{piano}.

\smallskip
\noindent{\bf Acknowledgments}\, \, This work is supported by the NSF of China under Grant No.11775306, No.11531004, No.11675113; the Fundamental Research Funds for the Central Universities Grants No.15CX05062A, No.16CX02049A; Simons Foundation No.523868; the Shandong Provincial Natural Science Foundation No.ZR2016AQ06; Qingdao applied basic research program No.15-9-1-103-jch, and a project sponsored by SRF for ROCS, SEM.

\end{document}